\documentclass[aps,twocolumn,groupedaddress,amsmath,superscriptaddress]{revtex4-1}
\usepackage{dcolumn}
\usepackage{bm}
\usepackage{amssymb}
\usepackage{txfonts}
\usepackage{braket}

\usepackage[pdftex]{hyperref}   
\usepackage{graphicx}
\usepackage{color}

\newcommand{\bk}{{\mathbf k}}

\newcommand{\br}{{\mathbf r}}
\newcommand{\bS}{{\mathbf S}}
\newcommand{\del}{{\partial}}
\newcommand{\eps}{{\varepsilon}}
\newcommand{\mM}{{\mathcal M}}

\newcommand{\mS}{{\mathcal S}}
\newcommand{\jsx}{ j^{s^z}_x} 
\newcommand{\jsy}{ j^{s^x}_y} 
 
\newcommand{\SIxx}{ \mS_{1}^{xx} } 
\newcommand{\SIIxx}{ \mS_{2}^{xx} } 
\newcommand{\SIyy}{ \mS_{1}^{yy} } 
\newcommand{\SIIyy}{ \mS_{2}^{yy} } 
\newcommand{\up}{\uparrow}
\newcommand{\down}{\downarrow}

\begin{document}
\title{Nonreciprocal spin Seebeck effect in antiferromagnets } 
\author{Rina Takashima}
\affiliation{Department of Applied Physics, The University of Tokyo, Tokyo 113-8656, Japan }
\email{r.takashima@aion.t.u-tokyo.ac.jp}
\author{Yuki Shiomi}
\affiliation{Department of Applied Physics, The University of Tokyo, Tokyo 113-8656, Japan }
\affiliation{RIKEN Center for Emergent Matter Science (CEMS), Wako 351-0198, Japan}
\author{Yukitoshi Motome}
\affiliation{Department of Applied Physics, The University of Tokyo, Tokyo 113-8656, Japan }
\date{\today}
\begin{abstract}
We theoretically propose a nonreciprocal spin Seebeck effect, i.e., nonreciprocal spin transport generated by a temperature gradient, in antiferromagnetic insulators with broken inversion symmetry. We find that nonreciprocity in antiferromagnets  has rich properties not expected in ferromagnets. In particular, we show that polar antiferromagnets, in which the crystal lacks the spatial inversion symmetry, exhibit perfect nonreciprocity --- one-way spin current flow irrespective of the direction of the temperature gradient. We also show that nonpolar  centrosymmetric crystals can exhibit nonreciprocity when a magnetic order breaks the inversion symmetry, and in this case, the  direction of the nonreciprocal flow can be controlled by reversing the magnetic domain. As their representatives, we calculate the nonreciprocal spin Seebeck voltages for the polar antiferromagnet $\alpha$-Cu$_2$V$_2$O$_7$ and the honeycomb antiferromagnet MnPS$_3$, while varying temperature and magnetic field. 
\end{abstract} 
\maketitle 
The reciprocal relation is a fundamental principle in thermodynamics assured by the symmetry of the system. It is, however, violated when a certain symmetry is broken, e.g., by crystal structures, electronic orderings, and external fields. Such violation of the reciprocity has attracted much interest from both fundamental physics and application. An archetype is the Faraday effect of light, in which the breaking of time-reversal symmetry causes a rotation of the polarization plane in an opposite direction when the propagation direction of light is switched. This nonreciprocal property has been used for an optical isolator and optical data storage. Another example is found in a $p$-$n$ junction, which allows a one-way flow of an electric current. A similar diode effect can also occur in a bulk crystal when time-reversal  and spatial-inversion symmetries are simultaneously broken~\cite{Rikken2001}. 

The nonreciprocity has also been studied for the propagation of spin waves in magnetic materials. The most pronounced example is the Damon-Eshbach mode, in which spin waves propagate on a material surface only in one direction~\cite{Damon1961}. Also in a bulk magnet, the breaking of spatial-inversion symmetry gives rise to nonreciprocal propagation of spin waves. There, an asymmetric exchange interaction called the Dzyaloshinskii-Moriya (DM) interaction~\cite{Dzyaloshinsky1958, Moriya1960} brings about asymmetry in the spin-wave dispersion with respect to the propagation direction. This has been experimentally observed in hetero-multilayer films of ferromagnets~\cite{Zakeri2010}, ferromagnets having noncentrosymmetric crystal symmetries~\cite{Iguchi2015, Seki2016,Sato2016a}, and a polar antiferromagnet (AFM) $\alpha$-Cu$_2$V$_2$O$_7$~\cite{Gitgeatpong2017a}. Since a spin wave can carry a spin current, the asymmetric dispersion may give rise to a nonreciprocal spin current. However, such an effect remains elusive thus far, despite the relevance to applications in spintronics as well as magnonic devices.

\begin{figure}[t]
\includegraphics[width=0.9\columnwidth]{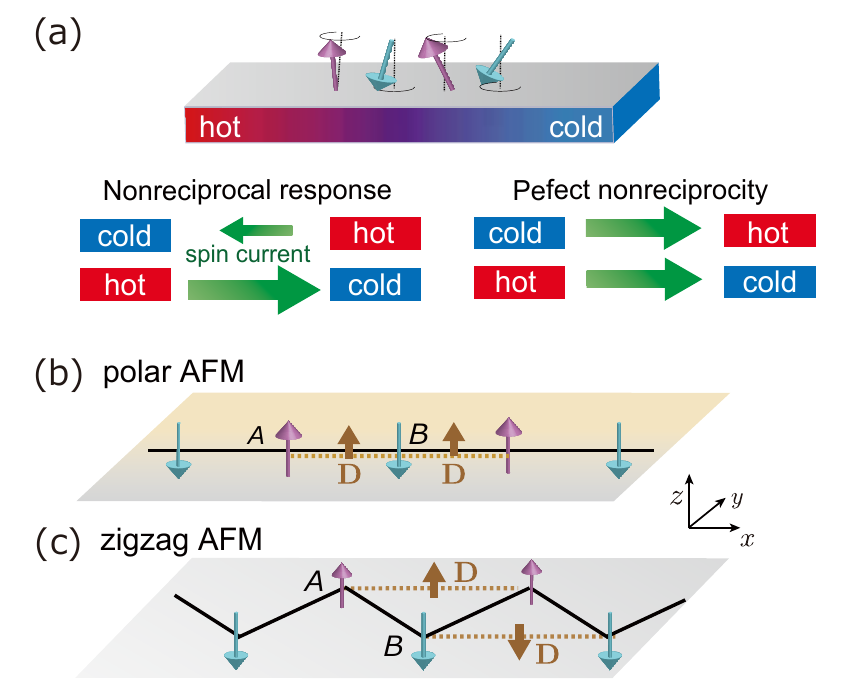}  
\caption{ 
(a) Schematic picture of a nonreciprocal spin current under a thermal gradient in an AFM. 
(b), (c) Schematic pictures of (b) a polar AFM  with a uniform DM interaction and (c) a zigzag AFM with a staggered DM interaction.  
The color gradient in (b) represents the breaking of mirror symmetry with respect to the $xz$ plane.
}
\label{fig:system}
\end{figure}

In this Rapid Communication, we propose a nonreciprocal response of a spin current in antiferromagnetic insulators, which stems from the asymmetric spin-wave dispersion.  
Specifically, we consider the spin Seebeck effect (SSE), a magnetothermal phenomenon in which a temperature gradient causes a spin voltage~\cite{Uchida2008b, Seki2015, Wu2016a, Shiomi2017a}. We show that a nonreciprocal spin current can be generated as a nonlinear response to a temperature gradient [Fig.~\ref{fig:system}(a)]. 
A nonreciprocal spin current response to an electric field was recently discussed in noncentrosymmetric metals~~\cite{Yu2014, Hamamoto2017}, which suffer from Joule heating. 
We here discuss the nonreciprocal SSE, mainly for antiferromagnetic insulators, as they have drawn considerable interest in recent spintronics owing to less stray field and ultrafast spin dynamics~\cite{Jungwirth2016, Baltz2018}. 
We find that the AFMs show remarkable properties in the nonreciprocal SSE, which are not expected in ferromagnets. 
We demonstrate that the nonreciprocal SSE appears in a different manner for two different types of AFMs: One is a polar AFM on a noncentrosymmetric lattice and the other is a zigzag AFM on a centrosymmetric lattice. 
The polar AFMs exhibit {\it perfect nonreciprocity}: A spin current flows only in one direction irrespective of the direction of the temperature gradient [Fig.~\ref{fig:system}(a)]. 
On the other hand, in the zigzag AFM, the nonreciprocity can be controlled by reversing magnetic domains.
For the experimental observations, we calculate the spin Seebeck voltages for candidate materials for the two cases, the polar AFM $\alpha$-Cu$_2$V$_2$O$_7$ and the honeycomb (two-dimensional zigzag) AFM MnPS$_3$, and clarify the dependence on temperature, the magnetic field, and the direction of the temperature gradient. 

We consider the spin current generated parallel to a temperature gradient up to the second order:  
\begin{eqnarray}
\jsx = \SIxx ( \del_x T)+\SIIxx (\del_x T)^2, \label{eq:SSE}
\end{eqnarray} 
where $\jsx$ is the  spin current that flows in the $x$ direction carrying the spin along the $z$ axis, and $T$ is the local temperature of the sample. 
 The first term in Eq.~\eqref{eq:SSE} corresponds to the conventional SSE~\cite{Uchida2008b, Seki2015, Wu2016a, Shiomi2017a}, and the second one is the nonlinear term, which we will discuss in this work. When $\SIIxx$ is nonzero, the magnitude of $\jsx$  changes depending on the sign of $\del_x T$. Thus, the nonlinear contribution in the SSE gives rise to a nonreciprocal spin current $\jsx$. 
We note that, from the symmetry point of view, such nonreciprocity is not allowed when the system is symmetric under the spatial inversion {$\mathcal I$} or mirror reflection with respect to the $xz$ plane, denoted by $\mM_y$. 
Meanwhile, the linear 
component $\SIxx$ vanishes when $\mM_x$ ($yz$ mirror) or $\mM_y$ symmetry exists. 

\begin{table}[t]
  \begin{tabular}{|
  c||ccc|c| c|c|} \hline 
&$\mathcal I$
& $\mM_x$ & $\mM_y$ &$B^z$ dep. of $\mS_1^{xx}$ \  & $B^z$ dep. of $\mS_2^{xx}$ & domain  \\ \hline
polar AFM & $\times$  
& $\checkmark$ & $\times$ & odd &even & indep.  \\ \hline  
zigzag AFM &$\times$   
& $\times$   &$\checkmark$   & odd & odd & dep. \\ \hline
\end{tabular}
\caption{
Symmetry arguments on the magnetic field dependence ($B^z$ dep.) of the SSE coefficients for the one-dimensional spin models for polar and zigzag AFMs [Eqs.~(\ref{eq:H_0})-(\ref{eq:DM_zigzag}); see also Figs.~\ref{fig:system}(b) and \ref{fig:system}(c)]. 
The domain dependence (dep.) or independence (indep.) of the nonreciprocal SSE on the magnetic domains are also shown. 
$\mathcal I$ is the inversion symmetry, and $\mM_x$ and $\mM_y$ are the mirror reflection symmetry with respect to the $yz$ and $zx$ planes, respectively. Note that each mirror symmetry represents the mirror reflection combined with half translation in the $x$ direction. 
 }
 \label{table:SSE}
\end{table}

In this study, we calculate the spin current in Eq.~(\ref{eq:SSE}) for AFMs, which exhibit richer nonreciprocal properties compared to ferromagnets~\footnote{See Supplemental Material at xxx for the SSE in a ferromagnets on  polar lattices and the derivation of Eqs. (9) and (10).}. We consider two types of noncentrosymmetric AFMs. One is an AFM on a noncentrosymmetric lattice, and the other is an AFM in which the inversion symmetry is broken by the magnetic order. As their typical examples, we first study one-dimensional spin models for the two types, which we call polar AFMs [Fig.~\ref{fig:system}(b)] and zigzag AFMs [Fig.~\ref{fig:system}(c)], respectively~\cite{Hayami2016a}. 
Their Hamiltonians are given by $H = H_{0} +H^{\rm polar/zigzag}_{\rm D} $, where 
\begin{eqnarray}
H_{0}=&  \sum_{r\neq r'}\left[ 
J_{rr'} \bS_{r} \cdot \bS_{r'} 
+ G_{rr'} (S^z_{r} S^z_{r'}- S^x_{r} S^x_{r'}-S^y_{r} S^y_{r'})\right] \nonumber\\
&\hspace{80pt} + g_s \frac{\mu_{B}}{\hbar}  
B^{z} \sum_{r} S_{r}^z.
\label{eq:H_0}
\end{eqnarray}
Here $\bS_{r} =(S^x_{r} , S^y_{r},S^z_{r})$ is the spin operator at $r=(i, \ell)$, where $i$ denotes  the unit cell and $\ell$ denotes the sublattice. 
We assume that a (magnetic) unit cell has two sites: $\ell=\{A, B\}$. $J_{r r'}$ and $G_{r r'}$ denote the coupling constants for the isotropic and anisotropic exchange interactions, respectively; the latter originates from the spin-orbit coupling. $g_s$ is the electron spin g-factor (we take $g_s=2$),  $\mu_B$ is the Bohr magneton, $\hbar$ is the reduced Planck constant, and $B^z$ is the magnetic field along the $z$ direction. $H^{\rm polar}_{\rm D}$ and $H^{\rm zigzag}_{\rm D}$ represent the DM interactions in the polar and zigzag  systems, respectively:
\begin{eqnarray}
&&H^{\rm polar}_{\rm D}= D \sum_i \hat{{\mathbf z}} \cdot \left(  \bS_{i, A} \times \bS_{i,B}  +\bS_{ i, B} \times \bS_{ i+1, A}  \right), 
\label{eq:DM_polar}\\ 
&&H^{\rm zigzag}_{\rm D}= D \sum_i \hat{{\mathbf z}} \cdot \left(  \bS_{i,A} \times \bS_{ i+1, A}  -\bS_{i, B} \times \bS_{ i+1, B}  \right), 
\label{eq:DM_zigzag}
\end{eqnarray}
where $\hat{\mathbf z}$ is the unit vector along the $z$ direction. 
Here, taking the chain direction as $x$, we assume that the polar system lacks $\mM_y$ symmetry while preserving $\mM_z$ symmetry ($xy$ mirror)
[Fig.~\ref{fig:system}(b)]; hence, we include a uniform DM interaction for all the nearest neighbors with the DM vector $\mathbf{D}\parallel \hat{\mathbf z}$ in Eq.~(\ref{eq:DM_polar}).  
On the other hand, in the zigzag system, there is no inversion symmetry at the centers of the second neighbor bonds,   
while the system is inversion symmetric with respect to the centers of the nearest neighbor bonds.  Therefore, we include a staggered DM interaction for the second neighbors with the DM vector $\mathbf{D}\parallel \hat{\mathbf z}$ in  Eq.~(\ref{eq:DM_zigzag}). 

Assuming a collinear antiferromagnetic ground state, namely, $\langle S^z_{i,A} \rangle = - \langle S^z_{i,B} \rangle = S$, we consider magnon excitations by using the Holstein-Primakoff transformation as 
\begin{eqnarray}
& S^+_{i, A} =\hbar (2S-a^\dag_{i}a_{i})^{1/2} a_{i}, \  
S^z_{ i, A} = \hbar\left( S- a^\dag_{i} a_{i} \right), \label{HP1}\\
& S^+_{ i, B} =\hbar b^\dag_{i} (2S-  b^\dag_{i} b_{i})^{1/2}, \ 
S^z_{i, B} = \hbar\left( b^\dag_{i} b_{i}- S \right), \label{HP2}
\end{eqnarray}
 where $S^+_{ i, \ell} =S^x_{ i, \ell}+ i S^y_{i, \ell} = (S^-_{ i, \ell})^\dag$. 
By substituting Eqs.~\eqref{HP1} and \eqref{HP2} into the Hamiltonian and using the linear spin wave approximation, we obtain the magnon Hamiltonian in the bilinear form of the operators of $a_i$ and $b_i$. Diagonalizing the Hamiltonian by the Bogoliubov transformation, we obtain
$H= \sum_{\sigma k_x}\eps_{\sigma k_x} \alpha^\dag_{ \sigma k_x} \alpha_{ \sigma k_x},  \label{eq:diag}$
where  $\alpha_{\sigma k_x}$ $(\alpha^\dag_{\sigma k_x})$ is the annihilation (creation) operator of a magnon with the spin angular momentum $\sigma=\{\uparrow, \downarrow\}$  and the momentum $k_x$. $\eps_{\sigma k_x} 
\geq 0 $ is the energy of the magnon. Because of the DM interaction, the magnon dispersion is deformed in an asymmetric manner with respect to $k_x$~\cite{Hayami2016a}. This is the crucial feature to produce the nonreciprocal SSE as discussed below. 

In the present systems,  the total spin along the $z$ direction $S_{\rm tot}^z \equiv\sum_{i} \left( S^z_{ i, A} +S^z_{ i, B} \right)  =\sum_{k_x} \left(-\alpha^\dag_{\downarrow k_x} \alpha_{\downarrow k_x}+ \alpha^\dag_{\uparrow k_x} \alpha_{\uparrow k_x} \right)$ is conserved, as the DM vectors point along the $z$ direction. Since each magnon excitation carries the spin angular momentum $\pm \hbar$, the local spin current density is given by 
\begin{eqnarray}
J^{s^z}_x= \hbar \int \frac{dk_x}{2\pi} 
\left( \varv^x_{\up k_x} n(\eps_{ \up k_x}) - \varv^x_{\down k_x}  n(\eps_{ \down k_x}) \right)
\label{eq:js}
\end{eqnarray}
where  the velocity is defined by $\varv^x_{ \sigma k_x} =(1/\hbar) \del \eps_{\sigma k_x}/\del k_x$, and $n(\eps_{\sigma k_x}) = \langle \alpha^\dag_{\sigma k_x}\alpha_{ \sigma k_x}\rangle $ denotes the magnon distribution at a finite temperature.   

To analyze the SSE, we use the Boltzmann transport theory~\cite{Rezende2016}. We assume that the temperature of the system has a linear gradient, $T(x)= T_0+\alpha x $,  where the coefficient $\alpha$ is small enough to allow us to define the equilibrium distribution of magnons by $ n^0(\eps_{ \sigma k_x}) = (\exp(\eps_{\sigma k_x}/T(x)) -1)^{-1} $. 
With the relaxation time approximation, the Boltzmann theory gives  
\begin{eqnarray}\varv^x_{\sigma  k_x}   \frac{\partial n(\eps_{ \sigma k_x})}{\partial x} =-\left. \frac{\del n}{\del t}\right|_{\rm col.}
=-\frac{n(\eps_{\sigma k_x})  - n^0(\eps_{\sigma k_x}) }{\tau}, 
\label{eq:Boltzmann}
\end{eqnarray}
where we have neglected the energy and momentum dependence of the relaxation time $\tau$. Substituting the solution of Eq.~(\ref{eq:Boltzmann}) into Eq.~(\ref{eq:js}) and 
averaging over the space
,  
we obtain the net component of spin current in Eq.~(\ref{eq:SSE}) with the coefficients of 
\begin{eqnarray}
\SIxx &=- \hbar\tau\int \frac{dk_x}{ 2\pi }\left[
(\varv^x_{\up k_x})^2 
f^{(1)}(\eps_{n\up k_x}) 
-(\varv^x_{\down k_x})^2 
f^{(1)}(\eps_{\down k_x}) 
\right],   \label{S1}\\
\SIIxx  &= \hbar\tau^2 \int \frac{dk_x}{2\pi  }
\left[  (\varv^x_{\up k_x})^3
f^{(2)}(\eps_{\up k_x}) 
- (\varv^x_{\down k_x})^3
f^{(2)}(\eps_{\down k_x}) 
\right] \label{S2},
\end{eqnarray}
where $f^{(1)}(\eps) =\del n^{0}/\del T|_{T=T_0}$ and $f^{(2)}(\eps) = \del^2 n^{0}/\del T^2|_{T=T_0}$~[18]. 
 Eq.~\eqref{S2} indicates that the nonlinear component originates in the asymmetry in the magnon dispersion, whose measure is given by the cube of the velocity averaged over a constant energy surface as \begin{eqnarray}\langle( \varv_{\sigma k_x}^{x})^3\rangle_{\eps_{k_x}=\eps} \coloneqq \int_{  \eps_{\sigma k_x}=\eps } \frac{dk_x}{2\pi }  ( \varv_{\sigma k_x}^{x})^3. \end{eqnarray}
This value vanishes when the magnon dispersion for each spin component is symmetric with respect to $k_x$.

As mentioned above, the polar and zigzag AFMs have the asymmetric magnon dispersions with $\langle ( \varv_{\sigma k_x}^{x})^3\rangle_{\eps_{k_x}=\eps}\neq 0$, and hence, they exhibit the nonreciprocal SSE. Due to the different symmetry, however, the SSE appears in a different manner between the two cases. 
As noted below Eq.~(\ref{eq:SSE}), the linear SSE coefficient $\SIxx$ can be nonzero only when both $\mM_x$ and $\mM_y$ symmetries are broken, whereas the nonlinear one $\SIIxx$ can be nonzero when $\mM_y$ symmetry is broken in addition to the inversion symmetry $\mathcal{I}$. In polar AFMs, where $\mM_y$ ($\mM_x$)  is (un)broken, 
$\SIxx$ vanishes but $\SIIxx$ may become nonzero at $B^z=0$. 
When the magnetic field $B^z$, which breaks $\mM_x$, is applied, $\SIxx$
is induced as an odd function of $B^{z}$, while $\SIIxx$ is an even function of $B^{z}$. 
On the other hand, in the zigzag AFMs, where $\mM_y$ is preserved, both $\SIxx$ and $\SIIxx$ are odd functions of $B^z$. The results are summarized in Table~\ref{table:SSE}.

From the symmetry arguments, an interesting phenomenon is readily concluded  for the polar AFMs. 
When $\SIIxx$ is nonzero at $B^z=0$ in a polar AFM, the SSE occurs even in the absence of the magnetic field. 
This is perfect nonreciprocal SSE, one-way flow of the spin current irrespective of the direction of the temperature gradient 
[Fig.~\ref{fig:system}(a)].

\begin{figure}[tb]
\includegraphics[width=0.95\columnwidth]{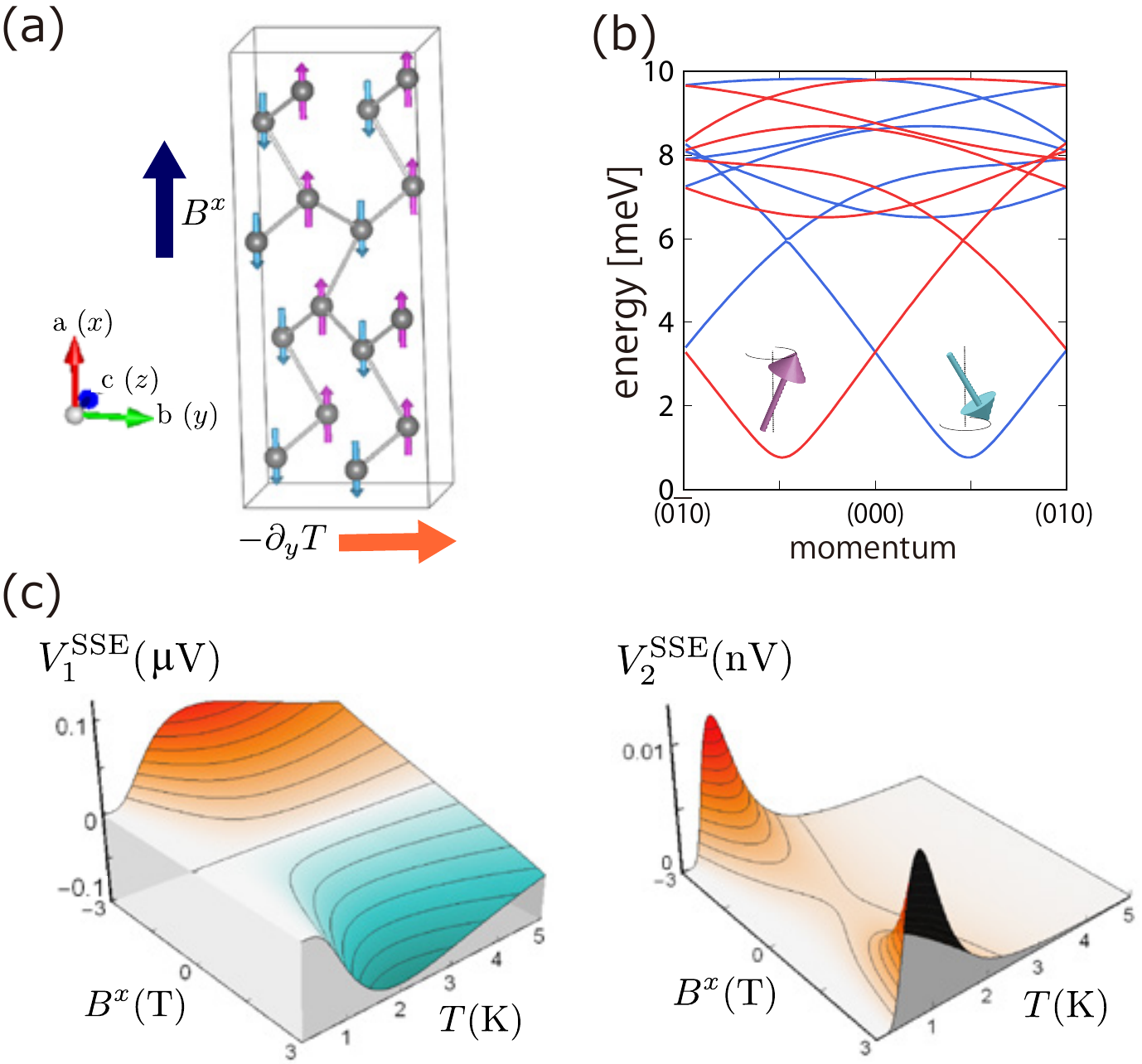}  
\caption{
(a) Schematic picture of the lattice structure and the spin configuration on the Cu$^{2+}$ ions in $\alpha$-Cu$_2$V$_2$O$_7$. The crystallographic axes are also shown. 
(b) Magnon bands in the polar AFM $\alpha$-Cu$_2$V$_2$O$_7$. The blue (red) bands carry the spin angular momentum $S_{\rm tot}^z=1(-1)$, and each band is asymmetric along $k_y$.  The parameters are given in the main text. 
(c) Dependence of the spin Seebeck voltages on the temperature and the applied field along [100]: the linear term $V_1^{\rm SSE}$ (left) and the nonlinear term $V_2^{\rm SSE}$ (right). 
}
\label{fig:CVO} 
\end{figure}　

Now let us estimate the coefficients given by Eqs.~(\ref{S1}) and (\ref{S2}) for real materials. First we consider a candidate for the polar AFMs,  
$\alpha$-Cu$_2$V$_2$O$_7$, whose lattice structure breaks the mirror symmetry with respect to the $ab$ plane [see Fig.~\ref{fig:CVO}(a)]. Below $T_N =33.4$~K,  
$\alpha$-Cu$_2$V$_2$O$_7$ shows an antiferromagnetic order, where Cu$^{2+}$ spins ($S=1/2$) align
 antiparallel  along $[100]$ [Fig.~2(a)] with a small canting along $[001]$~\cite{Gitgeatpong2015, Lee2016a, Gitgeatpong2017a, Gitgeatpong2017b}.  
The magnon bands obtained by a recent  neutron scattering experiment 
indicate the presence of the strong uniform DM interaction~\cite{Gitgeatpong2017a} similar to the polar AFMs discussed above.
In the following calculation, we use the model Hamiltonian, obtained via the inelastic neutron scattering experiment~\cite{Gitgeatpong2017a}. 
It has the isotropic exchange interactions between the first, second, and third neighbors, $J_1=2.67$, $J_2=2.99$, and $J_3=5.42$ in units of meV 
the nearest-neighbor anisotropic exchange interaction $G_1 =0.282$~meV, and the nearest-neighbor DM interaction $D=2.79$~meV; note that while the $(x, y, z)$ axes are taken along the crystallographic $(a, b, c)$ axes, they correspond to ($z,x,y$) in the model in Eqs.~(\ref{eq:H_0}) and (\ref{eq:DM_polar}) (the total spin angular momentum along the $x$ direction is conserved). 
Since each unit cell has 16 Cu$^{2+}$ spins, the magnon bands have eight branches per spin~\cite{Gitgeatpong2017a}, as reproduced in Fig.~2(b).  
The magnon dispersions are asymmetric along the $k_y$ direction resulting  in  $\langle( \varv_{n\sigma\bk}^{y})^3\rangle_{\eps_{n\sigma \bk}=\eps}\neq 0$ with $n$ being the band index. Hence, the system exhibits the SSE along the $y$ direction, $\jsy= \SIyy ( \del_y T)+\SIIyy (\del_y T)^2 $. 

In experiments, the spin current generated by the SSE can be measured by the inverse spin Hall effect of Pt attached to the sample. We assume that the induced voltage in Pt is simply given by the sum of linear and nonlinear components as $V^{\rm SSE}=V_1^{\rm SSE}+V_2^{\rm SSE}$, where
\begin{eqnarray} 
V_1^{\rm SSE}&= -\rho_{\rm Pt} \theta_{\rm sh}\frac{2e}{\hbar} L\SIyy (\del_y T) , \\
V_2^{\rm SSE} &=-\rho_{\rm Pt} \theta_{\rm sh}\frac{2e}{\hbar} L \SIIyy (\del_y T)^2,  
\end{eqnarray}
$\rho_{\rm Pt}$ is the electrical resistivity of Pt,  $\theta_{\rm sh}$ is the spin Hall angle of Pt, and $L$ is the length of the sample along the voltage direction.   
Recently, $V_1^{\rm SSE}$ was measured for $\alpha$-Cu$_2$V$_2$O$_7$, and $\tau\propto T^{-3}$ fits the experimental data well~\cite{Shiomi2017a}. 
In our analysis, assuming the power-law behavior, we estimate the magnitude of $\tau$ using the experimental data in Ref.~\cite{Shiomi2017a}. 
We use $\rho_{\rm Pt} =1.2\times 10^{-7}$~$\Omega$m  
and  $\theta_{\rm sh}=0.021$~\cite{Morota2011}, and set $\del_y T =10^3$~Km$^{-1}$ and $L=4\times 10^{-3}$~m based on the experimental setup~\cite{Shiomi2017a}.   
With the above assumptions,  we obtain the relaxation time $\tau \simeq c_0 /T^3$ with $c_0 =2 \times 10^{-9}$~K$^3$sec.

Using the obtained relaxation time, we calculate $V_1^{\rm SSE}$ and $V_2^{\rm SSE}$ as functions of the temperature and the field along the $x$ direction, $B^x$. 
The results are shown in Fig.~\ref{fig:CVO}(c). We find that the nonlinear component $V_2^{\rm SSE}$ appears as an even function of $B^x$, whereas the linear one $V_1^{\rm SSE}$ is odd.  
Furthermore, $V_2^{\rm SSE}$ is nonzero at $B^x=0$, i.e., the system exhibits the perfect nonreciprocal spin transport. 
These behaviors are exactly what we expected for the polar AFM; in the present material, instead of the mirror symmetry, the $C_2$ rotational symmetry along $[001]$ makes $\SIyy$ zero, while the breaking of  both inversion and $\mM_z$ symmetries results in nonzero $\SIIyy$. 

With regard to the temperature dependence, both $V_1^{\rm SSE}$ and $V_2^{\rm SSE}$ exhibit peaks at finite temperatures, and decay at higher temperatures, as shown in Fig.~\ref{fig:CVO}(c).  
Note that the calculated curve of $V_1^{\rm SSE}$ reproduces the experimental data well~\cite{Shiomi2017a}. 
The peak structure comes from the competition between the thermal excitations of magnons and the scattering rate.  
At a very low temperature, the SSE is enhanced by the thermal excitations of magnons as increasing temperature. With a further increase of temperature, however, the scattering processes, characterized by $\tau$, begin to suppress the SSE, leaving the peak structure at an intermediate temperature. 
We note that the peak temperatures are lower and the peaks are sharper for $V_2^{\rm SSE}$ compared to $V_1^{\rm SSE}$. 
This arises from the dependence on $\tau \propto T^{-3}$:  $V_1^{\rm SSE}$ and $V_2^{\rm SSE}$ 
depend on $\tau$ and $\tau^2$, respectively, as shown in Eqs.~(\ref{S1}) and (\ref{S2}). 

\begin{figure}[t!!] 
\includegraphics[width=0.95\columnwidth]{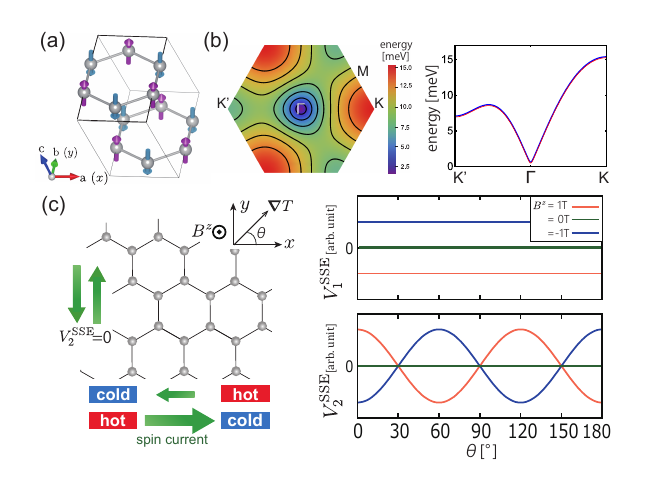}  
\caption{
(a) Schematic picture of the lattice structure and the spin configuration on the Mn$^{2+}$ ions in MnPS$_3$. Crystallographic axes are also shown. 
(b) Energy dispersion of the magnons. The energies of the magnons with $S_{\rm tot}^z=\pm 1$ are degenerate. 
(c) 
Dependence 
of the spin Seebeck voltages on the directions of the temperature gradient. 
The left figure represents a real-space picture of the directional dependence of the spin current. The magnetic field $B^z$ is normal to the honeycomb plane.
The right panels show the directional dependence of the linear term $V_1^{\rm SSE}$ (top) and the nonlinear term $V_2^{\rm SSE}$ (bottom). 
}
\label{fig:MPS} 
\end{figure}　

Next, we discuss a candidate for the zigzag AFMs, the honeycomb AFM MnPS$_3$. 
Note that the two-dimensional honeycomb structure is composed of one-dimensional zigzag chains running in three different directions. 
MnPS$_3$ has a layered honeycomb structure with the weak interlayer van der Waals interaction, as shown in Fig.~\ref{fig:MPS}(a). 
A neutron diffraction study shows that  Mn$^{2+}$ spins ($S=5/2$) align  in a staggered way below $T_{N} =78$~K, whose moment directions are almost normal to the honeycomb plane~\cite{Kurosawa1983} [Fig.~3(a)].  
Hereafter, we label the crystallographic coordinate $(a, b, c^*)$ by $(x, y, z)$, where $c^*$ is normal to the $ab$ plane.
The  spin model obtained by an inelastic neutron scattering~\cite{Wildes1998a} includes $J_1=1.54$, $J_2=0.14$, $J_3=0.36$, and $G_1=1.1 \times 10^{-3}$ in units of meV. 
Since the interlayer exchange interaction 
is much smaller than the intralayer exchange interactions, we calculate the SSE 
for a single honeycomb layer. 

From the lattice symmetry, the system has a staggered DM interaction between the second neighbors along the three types of zigzag chains, as in Eq.~(\ref{eq:DM_zigzag})~\footnote{
In a honeycomb lattice, it is allowed by the symmetry to have $D \sum_{\ell=\{1,2,3\}}\sum_i \bm z \cdot \left(  \bm S_{i,A} \times \bm S_{ i+a_\ell, A}  -\bm S_{i, B} \times \bm S_{ i+a_\ell, B}  \right)$, where $\bm r_{i+a_1} =\bm r_{i} +a \left(\frac{1}{2} ,\frac{\sqrt{3}}{2}\right)$,  $\bm r_{i+a_2} =\bm r_{i} + a(-1 ,0)$, and $\bm r_{i+a_3} =\bm r_{i} + a \left(\frac{1}{2} ,-\frac{\sqrt{3}}{2}\right)$.  Here $A$ and $B$ are the sublattice indices and $a$ is the length of the primitive vector.}. 
This leads to the asymmetry in the magnon bands, as shown in Fig.~\ref{fig:MPS}(b)~\cite{Hayami2016a}. This staggered DM interaction is reported to show  an interesting magnon transport, the ``Nernst'' effect of a magnon spin current~\cite{Zyuzin2016a, Cheng2016, Shiomi2017}.  In the experiment of the magnon Nernst effect~\cite{Shiomi2017},  the magnitude of $D$ has been estimated as $D \sim 0.3$~meV, which we adopt in the following analysis~\footnote{We note that the sign of $D$ has not been estimated.}.

In the above model, each unit cell has two sublattices  and the magnon bands with $S_{\rm tot}^z=\pm 1$ are degenerate. 
As shown in Fig.~3(b), the energy dispersion is asymmetric, namely, $\langle( \varv_{n\sigma\bk}^{x})^3\rangle_{\eps_{n, \sigma, \bk}=\eps}\neq 0$, e.g., along the K-$\Gamma$-K' line. 
In this situation, the nonreciprocal SSE appears 
when a magnetic field 
lifts the degeneracy of the two magnon bands ($S_{\rm tot}^z=\pm 1$). 
Similar to the one-dimensional zigzag AFM discussed above, both $V_1^{\rm SSE}$ and $V_2^{\rm SSE}$ are odd functions of the magnetic field normal to the honeycomb plane $B^z$.  
  
To experimentally detect the nonreciprocal SSE in this honeycomb system, we can exploit the directional dependence of $V_1^{\rm SSE}$ and $V_2^{\rm SSE}$. 
 Figure \ref{fig:MPS}(b) shows that the energy dispersion along the M-$\Gamma$-M cut is symmetric, which suggests that the nonreciprocal SSE does not occur along this direction.
Indeed, we find the directional dependence  with threefold rotational symmetry at nonzero temperature under a finite $B^z$, 
as shown in the lowerright panel of Fig.~\ref{fig:MPS}(c). (Note that the magnitudes depend on $\tau$, of which we do not have a quantitative estimate for the present compound.) 
The nonlinear spin Seebeck voltage $V_2^{\rm SSE}$ 
vanishes in the directions corresponding to M-$\Gamma$-M (e.g., $\theta=90^{\circ}$), whereas the linear one $V_1^{\rm SSE}$ (the upper panel) is always nonzero for $B^z\neq 0$ irrespective of the directions~\footnote{In our calculation within the linear spin wave theory, $V_1^{\rm SSE}$ does not show any angular dependence.}.

Finally, we make a remark on the controllability of the nonreciprocal spin current using the magnetic domain reversal.
In AFMs, there are energetically-degenerate magnetic domains connected by the time-reversal symmetry. 
As mentioned above, the breaking of inversion symmetry is necessary for nonzero $\SIIxx$. 
When the inversion symmetry is broken by a magnetic order as in the zigzag AFMs (e.g., MnPS$_3$),  $\SIIxx$ changes its sign between two different magnetic domains
 Therefore the nonreciprocal SSE can be controlled by reversing magnetic domains~\cite{Ressouche2010}. 
On the other hand, when the inversion symmetry is broken by the crystal structure as in the polar AFM {(e.g., $\alpha$-Cu$_2$V$_2$O$_7$), $\SIIxx$ is not changed by magnetic domain reversal. 
The results are shown in Table~\ref{table:SSE}. 

In summary, we have theoretically investigated the nonreciprocal  response of a spin current in AFMs under a thermal gradient. 
We showed that the nonreciprocal SSE appears in a different manner for the polar and zigzag AFMs. We found that the polar AFMs can exhibit perfect nonreciprocity, while the zigzag AFMs show a nonreciprocal SSE which can be controlled by reversing magnetic domains. 
For their experimental observations, we calculated the spin Seebeck voltage for $\alpha$-Cu$_2$V$_2$O$_7$ and the honeycomb antiferromagnet MnPS$_3$ while varying temperature and magnetic field. 
Our results would contribute to experimental observation of the nonreciprocal spin transport and to future application to spintronics devices.

We would like to thank M. Sato and T. J. Sato  for fruitful discussions. This research was supported by Grants-in-Aid for Scientific Research under Grants Nos. 15K05176, 17H04806, 18H04311, and 18H04215. R. T. is supported by a JSPS Postdoctoral Fellowship. 

\bibliographystyle{apsrev4-1}
%

\setcounter{equation}{0}
\setcounter{figure}{0}
\setcounter{table}{0}
\makeatletter
\renewcommand{\theequation}{S\arabic{equation}}
\renewcommand{\thefigure}{S\arabic{figure}}
\renewcommand{\bibnumfmt}[1]{[S#1]}
\renewcommand{\citenumfont}[1]{S#1}

\newpage
\widetext

\begin{center}
\vspace{1cm} 
\textbf{\large Supplemental Material for \\ \vspace{3mm} ``Nonreciprocal spin Seebeck effect in  antiferromagnets'' }
\end{center}
\section{Nonreciprocal spin Seebeck effect in a polar ferromagnet} 
Here we discuss the spin Seebeck effect in ferromagnets on polar lattices, which we call polar ferromagnets. We consider a one-dimensional model, whose Hamiltonian is given by $H = H^{\rm FM}_{0} +H^{\rm FM}_{\rm D} $, where 
\begin{align}
&H^{\rm FM}_{0}=  \sum_{i}\left[ 
J_{1} \bS_{i} \cdot \bS_{i+1}+ J_{2} \bS_{i} \cdot \bS_{i+2} 
+ G_{1} (S^z_{i} S^z_{i+1}- S^x_{i} S^x_{i+1}-S^y_{i} S^y_{i+1})\right]  
+ g_s \frac{\mu_{B}}{\hbar}  
B^{z} \sum_{i} S_{i}^z,\\
&H^{\rm FM}_{\rm D}= D \sum_i \hat{{\mathbf z}} \cdot \left(  \bS_{i} \times \bS_{i+1} \right).
\end{align}
Here $\bS_{i} =(S^x_{i} , S^y_{i},S^z_{i})$ is the spin operator at site $i$. 
$J_{1}$ ($J_{2}$) is the coupling constants for the isotropic exchange interactions for the first (second) neighbors, and $G_1$ is the anisotropic one for the first neighbors. $H^{\rm FM}_{\rm D}$ represents the uniform DM interaction for all the nearest neighbors with the DM vector $\mathbf{D}\parallel \hat{\mathbf z}$ as in the polar AFM in the main text. 
 
Assuming a collinear ferromagnetic ground state, namely, $\langle S^z_{i} \rangle = S$, we calculate the magnon dispersion $\eps_{k_x}$ within the linear spin wave approximation by using the Holstein-Primakoff transformation as 
\begin{align}
& S^+_{i} =\hbar (2S-a^\dag_{i}a_{i})^{1/2} a_{i}, \ \ \ 
S^z_{ i} = \hbar\left( S- a^\dag_{i} a_{i} \right), 
\end{align}
 where $S^+_{ i} =S^x_{ i}+ i S^y_{i} = (S^-_{i})^\dag$. 
Because of the DM interaction, the magnon dispersion becomes asymmetric with respect to $k_x$, which results in {a nonzero} $\langle ( \varv_{k_x}^{x} )^3 \rangle_{\eps_{k_x}=\eps}
$ {defined in Eq.~(11) in the main text}.  Noting that each magnon carries the spin angular momentum $-\hbar$, the local spin current density is given by \[ J^{s^z}_x=- \hbar \int \frac{dk_x}{2\pi} \varv^x_{ k_x}  n(\eps_{  k_x}),\] where  $\varv^x_{  k_x} =(1/\hbar) \del \eps_{ k_x}/\del k_x$ is the velocity  and  $n(\eps_{  k_x})$ is the distribution function of the magnon. 
Using the relaxation time approximation and averaging over the space, the coefficients in Eq.~(1) in the main text are given by  
\begin{eqnarray}
\SIxx &= \hbar\tau\int \frac{dk_x}{ 2\pi } 
(\varv^x_{ k_x})^2 
f^{(1)}(\eps_{ k_x}),  \\
\SIIxx  &=- \hbar\tau^2 \int \frac{dk_x}{2\pi  }
(\varv^x_{k_x})^3
f^{(2)}(\eps_{k_x}),
\end{eqnarray}
where the functions $f^{(1)}(\eps)$ and $f^{(2)}(\eps)$ are defined in the main text. 
In polar ferromagnets, the inversion, $\mM_x$, and $\mM_y$ symmetries are all broken. 
Therefore, both the linear coefficient $\SIxx$ and nonlinear one $\SIIxx$ can be nonzero even without a magnetic field, and 
they are even functions of $B^z$.

\section{Boltzmann theory for the spin Seebeck effect } 
In this section, we show the derivation of Eqs.~(9) and (10) from Eqs.~(1), (7), and (8) in the main text. 
In the small temperature gradient $T(x)= T_0+\alpha x $, 
where we assume that $T_0/\alpha$ is larger than the magnon mean free path, 
the equilibrium distribution function of magnons can be expanded as \[
 n^0_{\sigma \bk} (x)= \frac{1}{\exp(\eps_{\sigma \bk}/T(x)) -1}
\simeq \left. n^{0}_{\sigma \bk}\right|_{T=T_0}  + \alpha x \left .\frac{\partial  n^{0}_{\sigma \bk}}{\partial T}\right|_{T=T_0} + \frac{ \alpha^2 x^2}{2}\left .\frac{\partial^2  n^{0}_{\sigma \bk}}{\partial T^2}\right|_{T=T_0}{.} 
\]
{T}he distribution function $ n_{\sigma \bk} (\br)$ can also be expanded in the series of $\tau$ as 
\[
 n_{\sigma \bk} (x)\simeq
n^0_{\sigma \bk}(x) - \tau v^x_{\sigma k_x} \frac{\partial n^0_{\sigma \bk}(x)}{\partial x}  
+(\tau v^x_{\sigma k_x})^2 \frac{\partial^2 n^0_{\sigma \bk}(x)}{\partial x^2}.
\label{expand}
\]
Then the spin current in Eq.~(7) in the main text is given by 
\begin{align}
J^{s^z}_x &= \hbar \int \frac{dk_x}{2\pi} 
\left( \varv^x_{\up k_x} n(\eps_{ \up k_x}) - \varv^x_{\down k_x}  n(\eps_{ \down k_x}) \right)
\nonumber \\
&=-  \sum_{\sigma }   \hbar\tau \alpha    \int \frac{dk_x}{2\pi}  s_\sigma (v^{x}_{\sigma \bk})^2   \left .\frac{\partial  n^{0}_{\sigma \bk}}{\partial T}\right|_{T=T_0}
-  \sum_{\sigma }   \hbar \tau \alpha^2 x    \int \frac{dk_x}{2\pi}  s_\sigma (v^{x}_{\sigma \bk})^2   \left .\frac{\partial^2  n^{0}_{\sigma \bk}}{\partial T^2}\right|_{T=T_0}
+\sum_{\sigma}  \hbar \tau^2 \alpha^2\int \frac{dk_x}{2\pi} s_\sigma (v^x_{\sigma k_x})^3
 \left .\frac{\partial^2  n^{0}_{\sigma \bk}}{\partial T^2}\right|_{T=T_0},\label{spincu}
\end{align}
where $s_\uparrow(s_\downarrow ) =+1(-1)${; 
w}e have used that the following integration satisfies 
\begin{align}
  \int \frac{d k_x}{ 2\pi}   f(\eps_{\sigma \bk})v^{x}_{\sigma \bk} 
  =   \int \frac{d k_x}{2\pi}  f(\eps_{\sigma \bk}) \left( \frac{d \eps_{\sigma \bk}}{dk_x}\right)  = 
  \int \frac{d k_x}{ 2\pi} \frac{d }{d k_x} F(\eps_{\sigma \bk}) 
=\frac{1}{2\pi}\left[ F(\eps_{\sigma \bk}) \right]^{k_x=\pi}_{k_x=-\pi}=0, 
\end{align}
where $f(\eps_{\sigma \bk})$ is  an arbitrary function of an energy with which we can introduce $F(\eps)$ that satisfies $\frac{d}{d \eps} F(\eps)=f(\eps)$. 
Averaging $J^{s^z}_x$ in Eq.~(\ref{spincu}) over the space, we obtain Eq.~(1) with the coefficients Eqs.~(9) and (10) in the main text.

\end{document}